# Design, Environmental and Sustainability Constraints of new African Observatories:

## *The example of the Mozambique Radio Astronomy Observatory*

Domingos Barbosa, Miguel Bergano
Grupo de radioastronomia, Instituto de Telecomunicações
Campus Universitario de Santiago
Aveiro, Portugal

Valério A.R.M. Ribeiro
Astrophysics, Cosmology and Gravity Center, Department of Astronomy, University of Cape Town, Private Bag X3, Rondebosch 7701, South Africa

Anita Loots, Venkatasubramani L. Thondikulam
SKA-SA Project, NRF, 3rd Floor, The Park, Park Road, Pinelands 7405, South Africa

Michael Gaylard
Hartebeesthoek Radio Astronomy Observatory
PO Box 443 Krugersdorp 1740, South Africa

Arnold van Ardenne
ASTRON, Oude Hoogeveensedijk 4
7991 PD Dwingeloo
The Netherlands

Claudio Paulo[1,2]
[1]Dep. of Physics, University Eduardo Mondlane
Po Box 1569, Maputo, Mozambique
[2]School of Physics, University of the Witwatersrand
1 Jan Smuts Avenue, Braamfontein 2000
Johannesburg, South Africa

Sergio Colafrancesco
School of Physics, University of the Witwatersrand
1 Jan Smuts Avenue, Braamfontein 2000
Johannesburg, South Africa

José Carlos Amador[3,4]
[3]R&D Department, Martifer Solar SA
Zona Industrial, Ap. 17,
Oliveira de Frades, Portugal
[4]Rua da Imprensa, Nº 256, 6º Andar, Sala 601,
Maputo, Mozambique

Rodrigo Maia[*], Rui Melo[**]
Critical Software
[*]Pq. Industrial de Taveiro, Coimbra, Portugal
[**]Rua Pereira Marinho, 179, Maputo, Mozambique

*Abstract*—The Mozambique Radio Astronomy Observatory (MRAO) will be a first milestone towards development of radioastronomy in Mozambique. Development of MRAO will constitute a preparation step towards participation in the upcoming Africa VLBI Network and the Square Kilometer Array project. The MRAO first antenna is planned to serve as a capacitation and training facility and will be installed after the conversion of a 7-meter telecom dish in South Africa. Therefore, this first radiotelescope design has to comply with local spectral and environmental constraints. Furthermore, power availability and long term sustainability with potential inclusion of solar power and control of Radio Frequency Interference are analyzed. Here we outline some of the design, environmental and power sustainability constraints.

## I. OUT OF AFRICA: THE RADIOASTRONOMY LANDSACPE

Africa is embracing a scientific renaissance paving new avenues in cooperation towards radioastronomy infrastructures. Southern Africa's exquisite conditions for radioastronomy make this geography an excellent location for new world leading radioastronomical projects. Indeed, large scale projects like the African Very Large Baseline Interferometry Network (AVN) [1,2,3], the Square Kilometer Array (SKA) [4] and its precursor MeerKAT, and projects like PAPER [7] and CBASS [5] will contribute to change African scientific landscape through the creation of new world class astronomical observatories and digital support infrastructures.

### A. The African Very Large Baseline Interferometry Network

The AVN will use large radio telescopes across the continent, to be installed in the SKA African Partner countries – Botswana, Ghana, Kenya, Madagascar, Mauritius, Mozambique, Namibia and Zambia together with the operational VLBI station at Hartebeesthoek in South Africa. Where there are existing large satellite earth station antennas in these countries, their potential for economic conversion to VLBI-capable radio telescopes will be evaluated. In countries without large antennas, various options are possible, including developing smaller training telescopes from redundant small

The authors acknowledge URSI BEJ and Radionet (FP7) sponsoring.

antennas, and getting remote access to the large astronomy data sets that are available to develop an internal astronomy data reduction and science interpretation capability. The AVN will produce high-resolution observations of astronomical objects, contribute to geodetic measurements and monitor continental drifts through the accurate measure of the distance between each telescope in the network, and provide a serendipitous infrastructure for planetary space mission tracking in the Southern Hemisphere. The AVN will also play a key role in student training and capacity building preparing for the longer term endeavor of the SKA. The AVN is expected to contribute for the development of a common African radioastronomy language and aggregate local industries. While some of the trained fellows will constitute the core of a radioastronomy Research Area, most of the AVN and SKA trained scientists and engineers are expected to pursue later careers in other services and industry areas like ICT, enriching society and widening the socioeconomic impact of radioastronomy.

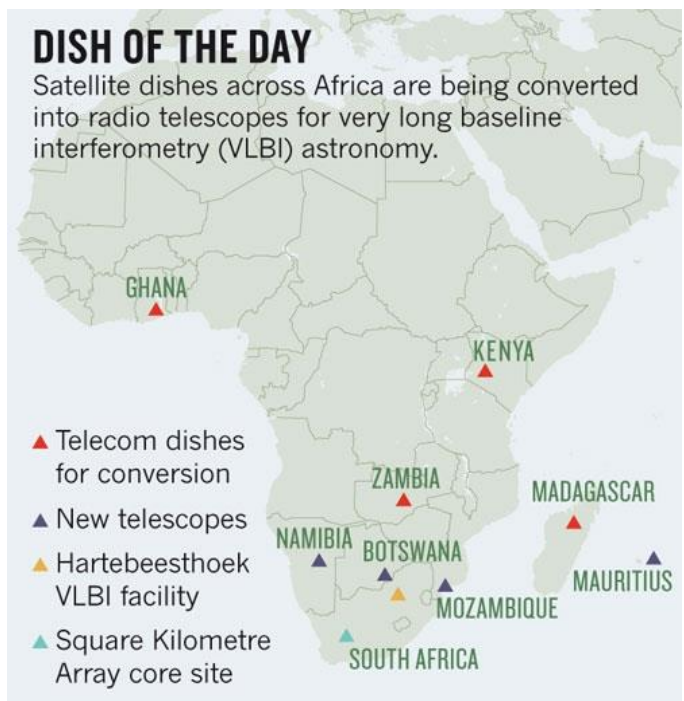

Fig. 1.– African planned radioastronomical landscape. From *SKA Africa*, @Nature [3]. This landscape will be further enriched with SKA Phase 2 extension plan, with the deployment of 2500 dishes and Aperture Array stations.

Starting with the South African Hartebeesthoek radiotelescope, already capable of VLBI services, the AVN will connect to European VLBI Network (EVN) across the pan-African UBUNTUnet Alliance and the pan-European GEANT digital infrastructures. The AVN will commence with the conversion of some large redundant telecommunications antennas into radio telescopes, like some 32-meter class satcom dishes (Ghana, Kenya, Zambia and Madagascar) with a later installation of new instruments and training facilities and potentially new 25-meter radiotelescopes in other countries.

### B. The Square Kilometer Array

The SKA [3] is an international Information and Computing Technology machine dedicated to radioastronomy that will be built in the Southern Hemisphere in high solar irradiated zones (South Africa with distant stations in the SKA African Partners - Botswana, Ghana, Kenya, Zambia, Madagascar, Mauritius, Mozambique, Namibia - and Australia/New Zealand). SKA, the only global project in the European Strategy Forum of Research Infrastructures, is a large-scale international science project involving 67 organizations in 20 countries, and counting with leading world industrial partners. SKA is a multipurpose radio interferometer with thousands of antennas linked together to provide a collecting area of one square kilometer. The SKA can be described a central core of ~200 Km diameter, with 3 spiral arms of cables connecting nodes of antennas spreading over sparse territories in several countries up to 3000Km distances. .

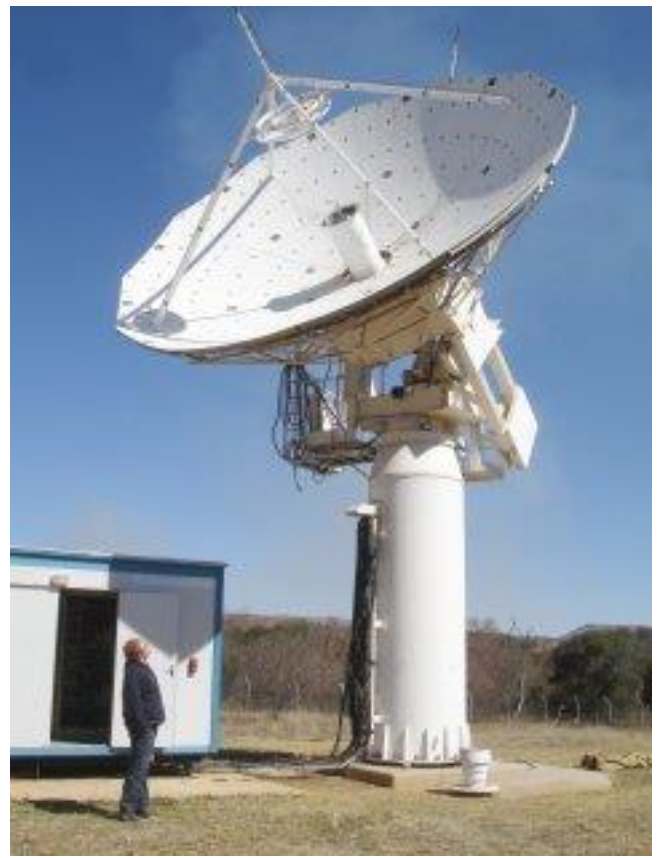

Fig. 2.– CBASS 7.6-meter radiotelescope, at the Hartebeesthoek observatory, after conversion from the original Telkom antenna. It is similar to the planned Mozambican 7.6-meter radiotelescope, currently in conversion from a sister Telkom antenna. From @CBASS.

It is expected SKA will start construction in 2017 spread over two phases with inclusion of new cutting-edge detection technologies enlarging Focal Plane Detector Arrays. Aperture Arrays (AA) stations are promised to enrich SKA Phase 2 and contribute to enhance the African SKA landscape [9,10] with an innovative multi-beam capacity capable of addressing the detailed mapping of the Universe's Dark Energy content and its properties. Also, it is expected that a substantial part of this project may become "green" during its lifetime, setting a pioneering example for self-sustainable mega-science production and infrastructure operation, with an expected direct economic and indirect societal impacts in the developing nations [6,8]. The two SKA sites (Southern Africa and Australia) were chosen for their exquisitely low Radio Frequency Interference (RFI), among other conditions, with the largest project extension to be located in Africa

II. THE MOZAMBIQUE RADIO ASTRONOMY OBSERVATORY

*A. Setting the Scene: preparing for large projects.*

With the AVN and later the SKA Phase 2 in the horizon, the SKA African Partner Countries launched Human Capital Development programs for the creation of a first community with the installation of facilities for hands-on training and technical capacitation. The International Year of Astronomy in 2009 (IYA 2009) by UNESCO already led to the onset of outreach astronomical activities in Mozambique, with seminars and observations in local schools, by local and international speakers about astronomy, exhibitions, teachers formation with the Galileo Teacher Training Program and led to explore cooperation between Mozambique and the International Astronomical Union through the Commission 46 "Teaching Astronomy for Development" [14].

The MRAO first antenna is being prepared in South Africa, in a collaboration of the Mozambique Government and the Department of Science and Technology (DST) of South Africa that donated the antenna, similar to the original antenna for the C-Band All Sky Survey (CBASS) project. These antennas were originally built by Intermediate Circular Orbit (ICO) for the South African Telkom Company and donated by this company to science. The South African DST has donated one of these antennas for conversion and installation as a pilot training facility in Mozambique. Therefore, the Mozambique antenna could gain from the lessons learned with the operational requirements and control block of the CBASS project and serve a first step towards technical capacitation.

The 7.6-meter Antenna for the establishment of MRAO is planned to be located at the Maluana site (lat: -25.4ºS, long: 32.64ºE), where the necessary infrastructure will be built by South Africa starting 2013 with foundations provided by Mozambique. The steps taken for Environmental Impact Assessment to ensure a long operational lifetime involve:

- Wind profile (check for cyclone history)
- Ground chemistry & acidity (attack on foundation's concrete), geotechnical characteristics of the site and geohydrology (especially depth of the water table)
- Soil Resistivity. This parameter is very important for earthing, evacuation of parasitic currents and design of the lightning protection
- Topography
- Control Infrastructure : Water, Power supply and Data keeping and control – 1Gps need for MRAO
- Ensure site and biodiversity preservation, mature trees for example.

Furthermore, spectral band allocation and protection will be discussed with the Mozambican radio spectrum regulator Instituto National de Comunicações de Moçambique (INCM) to ensure further protection of radioastronomical bands agreed under ITU regulations.

*B. Scientific and Technical Requirements*

Although such an antenna does not provide enough sensitivity for VLBI observations, most of the observations programs would be designed to cover relatively strong sources over extended period. Such an antenna could operate in 3 science operation modes in pointed observations:

- Radiometry with a multi-channel wideband radiometer
- Pulsar timing with a multi-channel wideband timing system
- Spectroscopy with a multi-channel narrow-band spectrometer

The Science modes condition the technical choices for the feed and receiver upgrade solution [11]. After careful weighting of the several options (maintain feed, or redesign a new subreflector) the feed/receiver solution adopted maintains the initial configuration with a retune of band filters. Hence, it would be possible to use all the available bandwidth of the S-band output for radiometry and pulsar timing; use the upper C-band retuned to include 6668 MHz for radiometry and spectroscopy; use the (circular) polarization outputs in each band to improve sensitivity. The frontend configuration uses uncooled Low Noise Amplifiers (LNA) with very good noise figures as a compromise between sensitivity and maintainability - avoid initial cost and complexity of a cryocooling solution. The digital backend will be based on a FPGA-based instrument such as a ROACH board [13]. A number of science cases to be explored in the two bands considered:

**S-Band:**

- Pulsar Timing: The strong Vela pulsar is visible from the South hemisphere. Although regular observations are possible with the neighboring HartRAO telescope, this telescope is highly demanded. A facility in Southern Africa monitoring regularly the Vela pulsar glitches, potentially up to 16 hours a day would be invaluable.
- Radiometry mode: follow the slow variation of strong radio sources like TauA and the bright quasar QSO 3C273.
- Radio Recombination Lines (RRL) are very difficult and complex to observe and would require additional work to allow spectroscopy.

**C-band:**

- Methanol Masers observations, after retune of filter for 6668.518 MHz. Monitoring of masers periods of

known periodic masers (range from about 29 to 509 days); Regional monitoring programs could be developed and conjugated with HartRAO and the new dishes in Southern Africa as part of AVN for daily monitoring

| Antenna Properties | |
|---|---|
| Primary Dish Diameter | 7.6 m |
| Azimuth sky coverage: | -270 ° to 270 ° |
| Elevation sky Coverage: | 0-90º |
| Operating Bands | S, C |
| half-power beamwidth at S-band | 1.36 ° (at 2000 MHz) |
| half-power beamwidth at upper C-band | 0.41 ° (at 6668MHz) |
| Drive speed: | Az: upto 10 deg/sec; El: upto 5 deg/sec |

Any observational program in such a wide band would require excision of potential RFI in either S or C bands, using multi-channel radiometry. This would constitute actually an important training aspect on spectrum management and monitoring.

### III. POWER SUSTAINABILITY: TOWARDS GREEN ?

The peak electrical requirements for the antenna, and related infrastructure containers are standard and are estimated to be 415Vac three phase 50Hz at 40kW. Any other electrical requirements and power infrastructure for future expansion of the facilities on site will be designed and added by the Mozambican team accordingly. Power stability must be ensured to control current peaks, for operation, cooling, computing and telescope management. Design of the power facility will require additional shielding to control and mitigate potential RFI from the electrical circuitry and ensure Electromagnetic Compatibility (EMC) of Power systems to avoid impairments to the radiotelescope sensitivity.

Although economic reasoning prefers a conventional power grid infrastructure to any radio observatory facility, to avoid power transport losses over large distances and keeping remote systems self-sufficient, solar power is being studied as a potential option for coupling to radiotelescopes. As a driver, it is expected that major projects like the SKA will integrate Green Energies into its power strategy during its lifetime and consider energy efficiency as part of a sustainable energy plan [8] influencing other infrastructures and facilities.

As a testbed for future developments, we consider options on Power availability through the use of Renewable Energy. Different solar power solutions are available, their choice depending on required power needs and again on economic reasoning. Design of any power facility for radioastronomical use will require additional shielding to control to mitigate potential Radio interference from the electrical circuitry. Since the first MRAO installations have initially relatively small power needs, we outline here a potential solution based on Solar Photovoltaic (PV) technology that could fit an off-grid solution. The main steps to design a PV based system have to include:

- Step1: Determine loads, location and inclination. Measure irradiation and check time series. Calculate installed power necessary to the load. Define number of modules and system specs (battery, charger, inverters).

- Step2: Choose worst irradiation month and determine its daily irradiation as Peak Sun hours.

- Step3: Define system configuration (Off-grid with battery or AC appliances with an inverter).

- Step4: Calculate installed power taking into account the system inefficiencies (cables, battery, regulator, and inverter) and choose Operation Voltage.

- Step5: Calculate number of modules in series.

- Step6: determine battery capacity $C_{bat}$, $C_{bat}=n$ (load/discharge depth), where $n$ is the numbers of days or period without solar charge.

- Step7: set regulator and inverter parameters ($V_{in}=V_{DC}$, $I_{in}=P_{peak}/V_{DC}$, $P_{out}$, $I_{out}$). In off-grid systems inverters is usually set to b 20% higher than the rated power from the added AC loads.

- Step8: Shield and measure EMC and RFI from components and subsystems. If possible use natural condition (hills, relief) to install and hide electrical production sub-systems.

- Use SmartGrid control software to ensure load adjustments and balancing and monitor the power system.

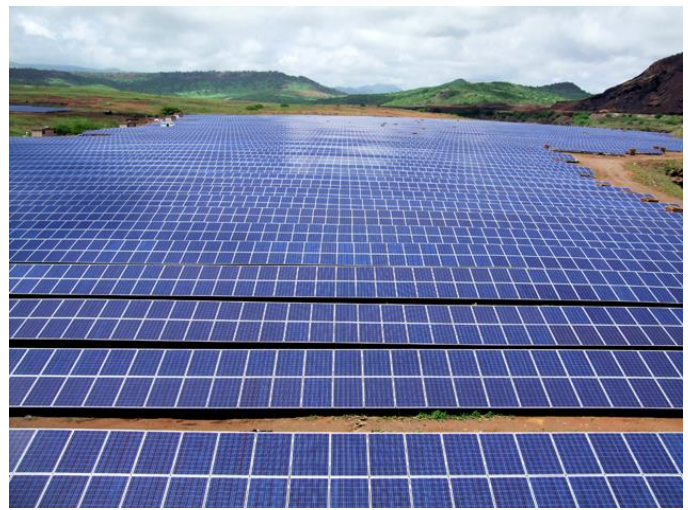

Fig. 3. An african solar initiative: Cape Verde PV plant [5MW] – Ground Fixed Structure. Courtesy Martifer.

### IV.

ACKNOWLEDGMENTS

The authors acknowledge fruitful discussions and suggestions by Prof. Miguel Avillez and his collaborator Gervasio Jorge Anela, both from U. Evora, Portugal that much improved the manuscript. VARMR acknowledges the South African SKA Project for funding the postdoctoral fellowship at the University of Cape Town. DB and MB acknowledge GEM grant support from IT and TICE. MB was supported by Fundação para a Ciência e Tecnologia (FCT) through grant SFRH/BD/76615/2011. The MRAO project is funded by the Department of International Relations and Cooperation (DIRCO), South Africa. The authors acknowledge support from the AERAP - African European Radioastronomy Platform.